# Effects of Starbursts on the Structure of Young Galaxies

Keiichi WADA

*Center for Information Processing Education, Hokkaido University, Sapporo 060, Japan*
*E-mail: wada@hipecs.hokudai.ac.jp*

Asao HABE

*Department of Physics, Hokkaido University, Sapporo 060, Japan*

*and*

Yoshiaki SOFUE

*Institute of Astronomy, University of Tokyo, Mitaka, Tokyo 181*

**Abstract**

We have investigated the effects of primeval starbursts on the galactic structure, showing that the dynamical structure of young galaxies is changed by relaxation of a stellar system formed by a starburst-driven outflowing gas (superwind). When a superwind interacts with halo gas, an expanding dense supershell of shocked gas is formed. Using a similarity solution for the expanding shell and a condition for the gravitational instability of a gaseous shell, we show that an expanding gaseous shell with a mass of several $10^{10} M_\odot$ and a radius of several kpc becomes gravitationally unstable to form stars. A stellar shell is thus formed, relaxes and will evolve into a fat stellar system.

In order to investigate the fate of the stellar shell and its dynamical influence on the host galaxy, we performed three dimensional $N$-body simulations of a stellar shell plus disk system. The evolution of the stellar shell was computed under two types of environment: an external flattened potential and a live stellar disk. We find that the relaxation process and resultant structure are significantly affected by the external disk potential, and also depend on the velocity dispersion of the initial shell stars and the mass of the shell. The final galactic structure ranges from a thick disk and a compact bulge with a high density core to a thin disk with a diffuse bulge. Isodensity maps of the computed disk plus bulge system resemble isophotes of S0 galaxies. We suggest that the effects of primeval starbursts would explain some observational properties of S0 galaxies.

**Key words:** Bulge – Starburst – S0 galaxies – Evolution of galaxies

## 1. Introduction

Observations have suggested that a substantial fraction of galaxies underwent starbursts (Soifer et al.1986). It is plausible that when the universe was younger, the number density of galaxies was high and that galaxy interaction was frequent. This would have triggered rapid gas fueling into galaxy nuclei, and starbursts would likely have occurred in a young and gas-rich disk galaxy. It is, therefore, worthwhile to investigate how primeval starbursts affected the evolution of structure and properties of young galaxies.

The formation of a galactic bulge in a disk galaxy is one of the long-standing questions in the evolution of galaxies. Larson (1976) showed that a bulge formed during an initial spherical contraction of a central core in a protogalactic gas sphere. This was followed by disk formation from a pancake contraction of the outer rotating part. This model has long been the basis for understanding bulge formation, although there have appeared several ideas that concern bulge formation after galaxy disks formed. Combes et al. (1990) showed that a bar-potential in the central region excites vertical motions of stars to form a boxy bulge; Pfenniger and Norman (1990) showed that contraction of disk stars toward the center in a barred potential results in an increase in velocity dispersion perpendicular to the disk, and yields a bulge-like distribution. However, in these studies effects of starbursts induced by galaxy-galaxy interactions in young galaxies were not considered. Recently, Sofue and Habe (1992) suggested formation of a central bulge by a primeval starburst, and showed that stars formed in starburst-driven gases would evolve into a bulge system, whose size depends on the strengths of starburst activity. However, their model is rather speculative and qualitative. They did not consider self-gravity of a newly formed stellar system and gravitational interaction of the bulge and disk system. In the bulge formation hypothesis by starbursts, these effects should be considered more carefully.

In the present paper, we intend to show that a substantial fraction of bulge stars could have originated in star formation in a starburst-driven supershell in primeval galaxies. In section 2, we evaluate properties of a stellar shell system by using a similarity solution of an expanding gas shell in an accreting ambient gas and a condition for gravitational instability. To investigate the evolution of this stellar system under the influence of the non-spherical external potential of a galactic disk, we perform three dimensional N-body simulations. Our methods and models are described in section 3. The models are chosen to show the effects of the initial conditions on the final structure and on the relaxation process. The initial conditions are within the limits provided by our analytical calculations. Numerical results are shown in section 4. Finally in section 5, we summarize our results and discuss implications of our results on the origin of S0 galaxies.

## 2. Gaseous and Stellar Shell Formation by a Superwind

### 2.1. Superwind Shell

Many hydrodynamical simulations have shown that a superwind produced by multiple supernovae interacts with the ambient gas and forms a hot cavity and a cooled dense gas shell (e.g. Tenorio-Tagle, Bodenheimer, Rozycka 1987; Tomisaka, Ikeuchi 1988; Mac Low, McCray 1989). Star formation in the gaseous shell would occur, if the shell is radiatively cooled and becomes gravitationally unstable (McCray, Kafatos 1987, Ostriker, Cowie 1981).

Recently, some high redshift radio-galaxies were found to have analogous properties to nearby far infrared galaxies (McCarthy, Persson, West 1992), in and around which intense star formation is ongoing. In these candidates of primeval galaxies, a superwind and a secondarily-formed stellar system along the radio-lobe axis have been observed (MacCarthy et al. 1987). It is suggested that an expanding super bubble in a gaseous halo triggers formation of a stellar system (Heckman, Armus, Miley 1990). Therefore, it seems reasonable to suppose that the stellar systems observed around high-redshift galaxies were formed by these processes.

We first estimate properties of a gaseous shell that would be formed by an interaction of a superwind with an accreting ambient gas. Since primeval starbursts and their interaction with proto-galactic gas sphere have not been fully studied, we here assume that they have similar properties to those observed in nearby active starburst galaxies.

From observations of double emission-lines with line splitting of 200−600 km/s in nearby far infrared galaxies, the presence of a "superwind" has been suggested with its "wind luminosity" of the order of $L_{\rm w} \geq 10^{12} L_\odot \approx 4 \times 10^{45}$ erg $s^{-1}$ (e.g. NGC 3690, IC 649, Arp 220, and NGC 6240), to $\sim 10^{13} L_\odot$ in IRAS 00182+7112 (Heckman, Armus & Miley 1990). Interaction of such a superwind with the ambient intergalactic/halo gas would produce a dense shell of shocked gas (superwind shell). It has been suggested that a stellar shell would be formed from such a dense gas shell, and that this could explain the origin of stellar shells around E/S0 galaxies (Umemura, Ikeuchi 1987; Williams, Christiansen 1985; Fabian, Nulsen, Stewart 1980), although mergers are the commonly accepted explanation of the shells and the ripples (e.g. Prieur 1990). A smaller scale but more massive stellar shell could form in the halo nearer to the nucleus.

Umemura and Ikeuchi (1987) presented a similarity solution of an expanding gaseous shell produced by a galactic wind interacting with an accreting ambient gas in a spherical galactic gravitational field. Their similarity solution of the expanding shell is given by $r_{\rm s}(t) = (at)^{2/3}$, when the initial density distribution of the accreting ambient gas is $\rho_{\rm g}(r) = \rho_{\rm c}(r/r_{\rm c})^{-1/2}$, where $\rho_{\rm c}$ and $r_{\rm c}$ are a central density and a core radius, respectively. The critical wind luminosity $L_{\rm w,crit}$, above which the wind can blow outward against the ram pressure of accreting gas, is given by

$$L_{\rm w,crit} = 8.9 \times 10^{42} \left(\frac{n_{\rm c}}{0.01\,{\rm cm}^{-3}}\right) \left(\frac{r_{\rm c}}{10\,{\rm kpc}}\right)^{\frac{1}{2}} \\ \times \left(\frac{M_{\rm g}}{10^{11}\,M_\odot}\right)^{\frac{3}{2}} \; {\rm erg}\; s^{-1}, \qquad (1)$$

when $a = 0.21({\rm kpc})^{-3/2}(10^6\,{\rm yr})^{-1}$, where $n_{\rm c}$ is the number density of the gas at the galactic center and $M_{\rm g}$ is a mass of the galaxy.

If the energy released from the starburst region were less than $L_{\rm w,crit}$, a super bubble confined in a dense gas shell with zero mean velocity would be formed (e.g. Tenorio-Tagle, Bodenheimer, Rozycka 1987; Tomisaka, Ikeuchi 1988; Mac Low, McCray 1989). It is most interesting when the ejected energy, $L_{\rm w}$, from a starburst region is comparable to $L_{\rm w,crit}$, because in this case the shell attains maximum radius and a large amount of ambient gas can be swept up into it.

In this case, we can estimate the shell radius $r_{\rm s}$, and mass $M_{\rm s}$, by using the similarity solution with the above condition for $a$, as

$$r_{\rm s}(t) = 7.8 \left(\frac{t}{10^8\,{\rm yr}}\right)^{\frac{2}{3}} \; {\rm kpc}, \qquad (2)$$

and

$$M_{\rm s}(t) = 1.2 \times 10^{10} \left(\frac{n_{\rm c}}{0.01\,{\rm cm}^{-3}}\right) \left(\frac{r_{\rm c}}{10\,{\rm kpc}}\right)^{\frac{1}{2}}$$

$$\times \left(\frac{t}{10^8 \, \text{yr}}\right)^{\frac{5}{3}} M_\odot . \tag{3}$$

## 2.2. Formation of a Stellar Shell

Next, we estimate a time when the expanding gas shell begins to fragment due to its self-gravity. Kinetic, gravitational, and thermal energy of a small part of the thin gas shell between an angle $\theta$ and $\theta + \Delta\theta$, the mass of which is $S(r_s)(r_s\Delta\theta)^2$, with $S(r_s)$ being the surface density of the shell at radius $r_s$, are defined as

$$E_k(t) \equiv \frac{1}{2} S(r_s)(r_s\Delta\theta)^2 (\dot{r}_s \Delta\theta)^2, \tag{4}$$

$$E_g(t) \equiv -\alpha \frac{G S(r_s)^2 (r_s\Delta\theta)^4}{r_s \Delta\theta}, \tag{5}$$

and

$$E_t(t) \equiv S(r_s)(r_s\Delta\theta)^2 \frac{kT_s}{\mu m_H (\gamma - 1)}, \tag{6}$$

respectively, where $\alpha$ is a constant of order unity, $m_H$ is the mass of a hydrogen atom, $\mu = 0.61$ is the mean molecular weight for a fully-ionized gas of $N_{\text{He}}/N_{\text{H}} = 0.1$, $\gamma = 5/3$ is the adiabatic exponent, and $T_s$ is a temperature of the gas shell. Using the adiabatic strong shock condition, the gas temperature $T_s(t)$ of the shell is

$$T_s(t) = \frac{3\mu m_H v_s(t)^2}{16k}, \tag{7}$$

where $v_s(t) \equiv \dot{r}_s(t)$. The total energy of the gas shell $E_{\text{total}} = E_k + E_g + E_t$ becomes less than zero, when

$$(\alpha G \Sigma r_s)^2 - 2\dot{r}_s^2 \frac{kT_s}{\mu m_H (\gamma - 1)} > 0. \tag{8}$$

Hence, equation (8) is a condition that the gas shell becomes gravitationally unstable.

Let us turn to a cooling time of the shocked gas at the leading edge of the expanding gas shell. From the energy conservation law for radiatively cooling monatomic gas with initial temperature $T_s$, the cooling time $t_{\text{cool}}$ of the shocked gas is given by

$$\begin{aligned} t_{\text{cool}} &= \frac{10 k^2 T_s^{2-\delta}}{3(2-\delta) A \rho_g(r_s) v_s^2} \\ &= 2.7 \times 10^3 \left(\frac{n_c}{0.01 \, \text{cm}^{-3}}\right)^{-1} \left(\frac{r_c}{10 \, \text{kpc}}\right)^{-\frac{1}{2}} \\ &\times \left(\frac{t}{10^8 \, \text{yr}}\right)^{-0.73} \text{yr}, \end{aligned} \tag{9}$$

where we assume the cooling function $\Lambda = AT^\delta$, with $\delta = -0.6$, and $A = 6.2 \times 10^{-19} \, \text{erg} \, \text{cm}^3 \, \text{s}^{-1}$ (McKee, Cowie 1977). Equation (9) shows that $t_{\text{cool}}$ is always less than the age of the shell, after $t = 2.3 \times 10^5 \, (n_c/0.01 \, \text{cm}^{-3})^{-0.58} (r_c/10 \, \text{kpc})^{-0.29}$ yr. Therefore, the temperature of the shell quickly cools to $\sim 10^4$ K at which it remains because of a secondary heating source: photoionization due to the nonthermal continuum emission which accompanies the wind as long as the nucleus remains active (Williams, Christiansen 1985).

The critical time, $t_{\text{crit}}$, when the gas shell just becomes gravitationally unstable, is derived from equations (2) and (8) with $T_s = 10^4$ K and $\alpha = 1$,

$$\begin{aligned} t_{\text{crit}} &= 1.6 \times 10^8 \\ &\times \left(\frac{n_c}{0.01 \, \text{cm}^{-3}}\right)^{-\frac{3}{4}} \left(\frac{r_c}{10 \, \text{kpc}}\right)^{-\frac{3}{8}} \text{yr}. \end{aligned} \tag{10}$$

This time scale is comparable to a typical starburst life time. The mass and the radius of the shell at $t_{\text{crit}}$ are

$$\begin{aligned} M_s(t_{\text{crit}}) &= 2.6 \times 10^{10} \\ &\times \left(\frac{n_c}{0.01 \, \text{cm}^{-3}}\right)^{-\frac{1}{4}} \left(\frac{r_c}{10 \, \text{kpc}}\right)^{-\frac{1}{8}} M_\odot, \end{aligned} \tag{11}$$

and

$$\begin{aligned} r_s(t_{\text{crit}}) &= 10.7 \\ &\times \left(\frac{n_c}{0.01 \, \text{cm}^{-3}}\right)^{-\frac{1}{2}} \left(\frac{r_c}{10 \, \text{kpc}}\right)^{-\frac{1}{4}} \text{kpc}, \end{aligned} \tag{12}$$

respectively. Soon after the starburst is over, the temperature of the shocked gas immediately cools to several $10^2$ K. At this stage the thickness of the gas shell is much less than its radius, because the internal pressure of the shell cannot support it against the external ram pressure.

Note that in above discussion we assumed that the cooling is very effective. However, if some heating mechanisms exist in the process, the temperature $T_s$ would remain higher than $10^4$ K and the critical time would be longer than that given by equation (10). Therefore in a higher density halo, a more massive gas shell than that estimated here would be expected.

At $t \sim t_{\text{crit}}$, fragmentation of the shell occurs due to gravitational instability, and star formation is triggered. In order to estimate mass of a stellar shell formed from the gas shell, we need the star formation efficiency $\epsilon_\star$. It has been suggested that $0.3 < \epsilon_\star < 0.5$ is needed to form bound star clusters from molecular clouds (Lada, Margulis, Derborn 1985). If we assume $\epsilon_\star$ in our primeval starburst shell was comparable to this Galactic value, a thin stellar shell of $M_{s,\text{star}} \sim$ several $\times 10^{10} M_\odot$ with $r_{s,\text{star}} \sim$ several kpc can be formed.

Further evolution of this stellar shell depends on the initial mass function of stars (IMF). If we assume that the IMF was similar to that obtained for the solar neighborhood (Miller, Scalo 1979), the birth rate of massive stars which would explode as SN can be estimated using the IMF: $\xi(\log M) \propto M^{-1.5}$, for $10 > M/M_\odot > 8$, and $M^{-2.3}$, for $M/M_\odot > 10$. This gives a rate of SN explosions as $4 \sim 10 \times 10^{-11}$ yr$^{-1}$ pc$^{-2}$, or $0.05 \sim 0.1$

yr$^{-1}$ in a shell of a radius 10 kpc. This rate would be maintained for the Jeans time of the gaseous shell, $\tau_{ff} \sim \sqrt{3\pi/32G\rho_{\text{shell}}} \sim 10^7$ yr ($\Delta r_s = 100$ pc, $M_s = 2.6 \times 10^{10} M_\odot$, $r_s = 10$ kpc). Hence, the number of SN which occurred in the shell is estimated to be $5 \sim 10 \times 10^5$, and the total energy release is about $10^{57}$ erg, which is comparable to the total self-gravitational energy of the gas shell, $GM_s^2/2r_s \sim 3 \times 10^{57}$ erg. This energy is not enough to destroy the gaseous shell, since about a few % of the energy released by SNs changes to kinetic energy of SN remnants (Spitzer 1978). Furthermore, if the fragmentation of proto-stellar cloud is hierarchical, there would be a sharper cutoff for the massive star formation. Larson (1982) has pointed out that massive stars are formed only in high-mass clouds. The Jeans mass in the gas shell is $M_J = (\pi kT/\mu G)^{3/2} \rho_{\text{shell}}^{-1/2} \sim 4 \times 10^4 M_\odot$ for $T = 100$ K. According to Ashman (1990), the largest fragment in a proto-stellar cloud, whose density fluctuation is $\xi(m) \propto m^{B-1}$, is given by

$$m_{\max} \sim \left(\frac{B}{B+1}\right)^{\frac{1}{B}} \left(\frac{m_{\min}}{M_\odot}\right)^{\frac{B+1}{B}} \times \left(\frac{M_c}{M_\odot}\right)^{-\frac{1}{B}} M_\odot. \quad (13)$$

Using $M_c = M_J$, $B = -2$, and taking $m_{\min} = 0.004\, M_\odot$ (Palla, Salpeter and Stahler 1983), we obtain $m_{\max} = 9\, M_\odot$. For $B = -3$, we obtain $m_{\max} = 0.75\, M_\odot$. Silk (1977) suggested $B = -2.9$ for $M > 0.6\, M_\odot$, if dynamical dissipation caused by turbulent motion in the proto-stellar cloud is effective, which would be likely excited by a proto-stellar wind as well as by a hydrodynamical instability in the shell. Therefore, it is probable that the IMF in the gaseous shell would have a sharp cutoff at around $\sim 1\, M_\odot$, and low mass stars dominate in the shell.

## 3. Models and Numerical Method

Suppose that a massive stellar shell comprising small-mass stars is formed. Then, our next interest is to know how the stellar shell and the disk system evolve. The stellar shell would finally relax and evolve into a spherical system distributed around the nucleus. We may thus expect that the system would evolve into a bulge-like structure. In order to investigate how the stellar shell would relax, and what shape and kinematic of the final "bulge" would exhibit, we perform three-dimensional $N$-body simulations.

### 3.1. Model of stellar shells

The parameters for the stellar shell can only be roughly estimated by the preceding analysis. We therefore choose a variety of allowed initial conditions, and concentrate ourselves on how they affect the final fate of the shell.

Table 1.. Parameters of the fixed potential models

| Model | $M_s/M_d$ | $\sigma_s$ | RESULT Figure | Type |
|---|---|---|---|---|
| $A_1$ | 0.5 | 10.0 | — | III |
| $A_2$ | 0.5 | 5.0 | Fig. 3 | III |
| $A_3$ | 0.5 | 1.0 | — | II |
| $A_4$ | 0.5 | 0.0 | — | II |
| B | 0.1 | 1.0 | Fig. 2 | II |
| C | 0.01 | 1.0 | Fig. 1 | I |
| D | 0.0 | 1.0 | — | I |

$M_s/M_d$: the ratio of the mass of the stellar shell and the disk mass. $\sigma_s$: initial velocity dispersion of the shell stars. Model D is a test particle model. Type I, II, and III mean relaxation types defined in Section 4.1. See also Fig. 4.

We assume that the initial stellar shell is spherically symmetric and thin ($\Delta r_s/r_s = 0.05$), and has zero mean velocity (i.e. no rotation and no radial motion). To model a stellar shell, 5000 collisionless particles are randomly distributed on a spherical shell. The main parameters of the models are the mass of the shell, $M_s$, and the initial velocity dispersion of the shell stars, $\sigma_s$. We simulate the evolution of the stellar shells in two types of model disk: (a) a fixed (i.e. time-independent) axisymmetric flat potential, and (b) a live thin stellar disk modeled by $N$-body selfgravitating particles.

### 3.2. Fixed potential models

The fixed disk potential is assumed to be the Miyamoto-Nagai potential (Miyamoto & Nagai 1975):

$$\Phi_{\text{MN}}(R,z) = -\frac{GM_d}{\sqrt{R^2 + (a + \sqrt{z^2 + b^2})^2}}, \quad (14)$$

where $a = 20$ kpc, $b = 2$ kpc, and $M_d = 1.0 \times 10^{11} M_\odot$. In this paper we use the following units: $[L] =$ kpc, $[M] = M_\odot$, $[T] = 10^8$ yr, and $[V] = [L/T] \sim 10$ km s$^{-1}$. The initial radius of the shells is 5.0.

In Table 1 we summarize the parameters of the initial shells in the fixed potential models. Model D is a test-particle model in which we do not take into account the self-gravity of the shell.

Since the model A series has a large $M_s/M_d$, the assumption of a fixed disk potential may be flawed. However, as we describe in section 4, the evolution of these massive shells is similar in both the fixed and lived potential.

Table 2.. Parameters of the live disk models

| Model | $N_d$ | $N_s$ | $M_s/M_d$ | $\sigma_s$ | Figure |
|---|---|---|---|---|---|
| S1 | 10000 | 2000 | 0.2 | 1.0 | — |
| S2 | 10000 | 5000 | 0.5 | 1.0 | Fig. 5 |
| S3 | 10000 | 5000 | 0.5 | 10.0 | Fig. 6 |

$N_d$ and $N_s$ are the number of the disk stars and of shell stars, respectively.
Shell of model S2 and that of S3 correspond to $A_3$ and $A_1$ of the fixed potential model, respectively.

### 3.3. Live disk models

The second group of models is calculated in order to investigate gravitational interaction between the shell and disk stars. Both the shell and disk are represented by collisionless particles (Table 2). The initial density profile of the disk is

$$\rho_d(R,z) = \frac{\Sigma_0}{\sqrt{2\pi}h} \exp\left(-\frac{R}{R_0} - \frac{z^2}{h^2}\right), \quad (15)$$

where $\Sigma_0$ is a surface density at $R = 0$, and $R_0$ and $h$ are the radial and disk scale lengths, respectively.

The vertical scale height $h$ and the vertical velocity dispersion $\sigma_z$ are determined so that the disk is quasi static in the vertical direction. The radial velocity dispersion at $R = R_0$ is determined from $\sigma_R(R_0) = 3.36 Q G \Sigma_0/\kappa(R_0)$, where $Q$ is Toomre's disk stability parameter (Toomre 1963), and we take $Q = 1.2$. The vertical velocity dispersion $\sigma_z$ is assumed to be $\sigma_z = 0.5\sigma_R$. The scale height of the disk is assumed to be constant in the disk, and is derived from the hydrostatic condition and the Poisson's equation of a very flattened system,

$$\frac{1}{\rho_d}\frac{\partial \rho_d \sigma_z^2}{\partial z} = -\frac{\partial \Phi}{\partial z}, \quad (16)$$

and

$$\frac{\partial^2 \Phi}{\partial z^2} = 4\pi G \rho_d + \frac{1}{R}\frac{\partial R F_R}{\partial R}, \quad (17)$$

where $F_R$ is a radial force of gravity of the disk (Binney, Tremaine 1987).

We also adopt a fixed dark halo potential: $\Phi_{\rm dark} = -GM_{\rm dh}/\sqrt{r^2 + r_0^2}$, where $M_{\rm dh}$ and $r_0$ are the mass and core radius of the dark halo. We take $r_0 = 10$ and $M_{\rm dh} = 5.0 M_d = 5.0 \times 10^{11} M_\odot$ so as to satisfy the criterion of a stable disk against the bar instability (Ostriker and Peebles 1973). The initial radius of the shells in these models is 3.0.

After several rotations of the disk, a shell is added to the disk, and we calculate the time evolution of the shell plus disk system.

### 3.4. Numerical Method

$N$-body calculations were performed on a workstation with GRAPE-3. The basic equations of the calculations are given by

$$m_i \frac{d^2 \boldsymbol{x}_i}{dt^2} = \boldsymbol{f}_i \quad (18)$$

and

$$\boldsymbol{f}_i = -G \sum_{j \neq i} \frac{m_i m_j (\boldsymbol{x}_j - \boldsymbol{x}_i)}{(r_{ij}^2 + \epsilon^2)^{3/2}}, \quad (19)$$

where $\boldsymbol{x}_i$ and $m_i$ are the position and the mass of the $i$-th particle, $\boldsymbol{f}_i$ is the force exerted on it, $r_{ij}$ is the distance between the $i$-th and $j$-th particles, $G$ is the gravitational constant, and $\epsilon$ is a softening parameter. We adopt a constant $\epsilon = 0.08$ and all particles have the same mass $m_i = m$ in our simulations. Instead of adopting methods to reduce the $N^2$ operations such as the particle-mesh method with the Fast Fourier Transform or tree algorithms, the force calculation was performed directly on GRAPE.

GRAPE (GRAvity PipE) is a series of special-purpose computers to accelerate solving the above equations, developed at the Department of Earth Science and Astronomy, University of Tokyo (Sugimoto et al. 1990). GRAPE-3 is a parallel-pipeline system, which attains 15 Gflops at peak speed (Okumura at al. 1991). Error analysis of the GRAPE was done by Makino, Ito and Ebisuzaki (1990): Integration error of the total energy per crossing time is about 0.01 percent for a stable Plummer sphere using $10^3$ particles. In our calculations the energy is conserved to within 1 percent per dynamical time. We also performed a number of comparison simulations by another numerical method (Particle-Mesh method with FFT), and found the same results as those obtained using GRAPE.

All other calculations, such as the time integration, were performed on the host computer (workstation). The time integration was performed using a standard leap-frog method.

### 4. Numerical Results

#### 4.1. Fixed potential models

The *violent relaxation* for collisionless N-body systems with no disk potential is governed by the initial ratio of kinetic to potential energy (e.g. van Albada 1982). Relaxation processes of 'cold models' (where this ratio is low) is more violent than that of 'hot models'. By adding a disk potential, the variety of relaxation processes is increased, with $\sigma_s$ a key determinant. We found that the relaxation process of seven models shown in Table 1 can be classified into three types; Type I (model C and D), Type II ($A_3$, $A_4$, and B), and Type III ($A_1$ and $A_2$).

The final structure and amount of interaction with the disk vary by type. Type III models have larger $\sigma_s$ than Type I or Type II. Type I models are dynamically hotter than Type II.

Figure 1 shows the time evolution of the positions of $N$-body particles projected onto the x-z plane (edge-on view) for model C. Due to the disk potential, stars near the disk fall toward the disk plane. A dense ring begins to form ($t = 0.68$) and expands in the z- direction ($t = 1.28$). At $t = 1.85$ a similar ring begins to form in a cigar-shaped shell at the disk plane. These processes are repeated and, as a result, a multiple-cylinder structure is formed ($t = 3.67 \sim 5.37$). The shell stars are finally relaxed to form a vague bulge ($t = 11.2$). The same structure is observed even in the non-selfgravitating model D. The multiple-cylinder structure is, therefore, a pattern caused by phase differences between particles oscillated in a non-spherical external potential; phase-mixing relaxes the system in model C, as in model D. The self-gravity of the shell is not important for the relaxation process in Type I model. The time needed for the shell stars to be relaxed in Type I models is more than 10.

Next, we show a typical example of a Type II model, in which shells are more massive and dynamically colder than in Type I models (Fig. 2; model B). The relaxation process and the final structure are very different from those of Type I. In the early stage, gravitational instability occurs in the collapsing shell. After the collapse, the shell expands non-isotropically, and forms a torus ($t = 0.77 \sim 1.15$). Gravitational instability grows in the torus, and two clumps are formed in the disk plane ($t = 1.67$). The clumps finally merge at $t \sim 3.5$ and form a compact core with a diffuse halo ($t = 8.76$). The effect of the disk potential is also important for the relaxation process in Type II models, although the self-gravity dominates the process.

Evolution of models with velocity dispersion larger than that in Type II is classified as Type III. The cylindrical structure no longer appears in these models (Fig.3). This shows that the disk potential does not affect the relaxation process so much. In model $A_1$, a shell is relaxed within $\sim 2 \times 10^8$ yr, and a diffuse prolate bulge is formed along the disk plane.

The square of the initial velocity dispersion $\sigma_s$ of each model is plotted against $M_s/M_d$ in Fig. 4. Three evolutional type in our results can be explained, if there are a critical value of $\sigma_s$ and $M_s/M_d$. For large $M_s/M_d$, the shell evolves as if there is no disk potential, provided that $\sigma_s$ is small enough. This is categorized as Type II, and selfgravity of the shell dominates the evolution. When $M_s/M_d$ and $\sigma_s$ is small, the shell is strongly affected by the disk potential and $M_s/M_d \sim 0.03$ is a critical ratio. In a non-spherical potential, a phase-mixing occurs during the shell relaxation near the disk plane. The vertical velocity dispersion ($\sigma_d$) of the disk stars in the Miyamoto-Nagai potential (eq. (14) to (16)), is critical for the phase-mixing by the disk potential to occur. This velocity dispersion is shown in Fig. 4 for two radii, $R = 1.0$ and $R = 10.0$. If the initial velocity dispersion of the shell stars is far greater than $\sigma_d$, the disk potential hardly changes the random motion of shell stars, that is, relaxation of the shell stars is not influenced by the disk potential. On the other hand, if the velocity dispersion is smaller than $\sigma_d$, the disk potential causes phase-mixing of shell stars. Thus evolution process in a disk potential depends on the initial velocity dispersion of the shell stars ($\sigma_s$) as well as the mass of the shell ($M_s$), therefore three types of evolution processes appear in our models.

The final structure of the remnant of the shell stars (bulge) is different among types I, II, or III. If the mass of the shell is very small, as in Type I, the bulge has a very vague shape (Fig. 1). On the other hand, the bulge in Type II (Fig. 2) or in Type III (Fig. 3) has a core-halo structure, although the core of Type II is smaller than that of Type III. The density profile of the bulge is $\rho(r) \propto r^{-2}$ at small $r$ and $\rho(r) \propto r^{-4}$ in the outer region. This profile is characteristic in remnant formed by the *violent relaxation* of a collisionless self-gravitating $N$-body system (e.g. van Albada 1982).

### 4.2. Live disk models

Concerning the shell evolution, results of models S1, S2 and S3 are similar to those of the fixed potential case.

We can see the following points from Fig.5, which shows evolution of (a) disk stars, (b) shell stars and (c) both disk and shell stars, in model S2, in which $M_s/M_d$ and $\sigma_s$ are equal to those in model $A_3$. Shell stars evolve in almost the same way as in Type II of the fixed potential models; After central concentration, the system of shell stars expands and changes into a cylindrical shape. In this torus two clumps grow early ($t = 0.21$) near to the disk plane. These two clumps finally merge and settle into a dense core with a diffuse halo at $t \sim 0.5$.

Disk stars are strongly influenced by the shell relaxation; stars in the central region follow a similar evolution to the shell stars: a central dense core is formed at the central concentration stage of the shell stars, and a clumpy structure is also formed at the same position as the shell stars ($t = 0.21$ and $0.31$). The disk stars are heated up in these processes, and the stellar disk rapidly thickens during the relaxation of the shell (see below). Moreover it is found that an oval ring density wave expands in the disk stars($t = 0.21 \sim 0.5$). As a result, the spiral arms seen in the initial disk disappear quickly.

Evolution of model S1, in which the mass of the shell is smaller than that of S2, is similar to the results of model S2, although the density wave formed in the disk is weak. Model S3, in which the shell has same $M_s/M_d$ as in model $A_1$, evolves differently from model S2 (Fig.

6). The density wave generated in the disk is not oval and the shell stars do not form clumps as seen in Type II models and in S2 model.

Figure 7 shows the time variation of an average scale height of the stellar disk in models S1, S2, S3, and NS. Model NS is for comparison, in which a spherical fixed potential $\Phi = GM_{\rm NS}/\sqrt{r^2 + r_{\rm c}^2}$ is added to the live disk instead of a stellar shell. The core radius $r_{\rm c} = 0.3$ and $M_{\rm NS}$ is determined so that the potential energy within $r_{\rm c}$ equals that of the initial shell of model S2. Figure 7 shows that a thick disk is formed in models with shells which are massive and have a small initial velocity dispersion.

Figure 8 shows the time evolution of mean vertical velocity dispersion of the whole disk in model S2, and S3. As with the scale height (Fig.7), the velocity dispersion increases rapidly for S2 in an early time. In model S2 the vertical velocity dispersion in the disk increases most rapidly and the disk stars are heated up, where the dense clumps are formed (cf. Fig.5). Since these clumps finally merge at the central region of the disk, the *hot* regions also merge, and a region with higher velocity dispersion is formed. This results in an increase in the vertical velocity dispersion and therefore in the scale height in the inner region of the disk. The increase of velocity dispersion is caused by a non-linear growth of the perturbation in disk stars driven by the evolution of shell stars. Due to this non-linear growth, potential energy of the region increases, and this causes the heating. Since hot shells do not produce clumps in a disk plane, it is reasonable that heating up of disk stars is more effective for a dynamically cold shell, i.e. S2 models (Type II ) than for S3 (Type III ).

In Fig.9, we plot time variation of the angular momentum of the shell and disk: About 1.2 % (S1), 5.8 % (S2), and 3.0 % (S3) of the total disk angular momentum is transferred from the disk to the shell during relaxation. Due to this angular momentum transfer, the bulge formed from the non-rotating shell begins to rotate in the same direction of the disk. The mean rotational velocity of the bulge is about 5 at $R = 1$ in model S2. The angular momentum transfer is caused by dynamical friction between the clumps and disk stars. When the shell is dynamically cold, such as that in model S2, a greater number of shell stars evolve into clumps in the disk plane (Fig.5). In fact, the angular momentum transfer is found to be highest in model S2.

The bulges of model S2 and S3 correspond to that of Type II and Type III , respectively. Figure 10 shows final radial density profiles of the shell stars for model S2 and S3. The bulge of both models have a core ($\rho(r) \propto r^{-2}$) and halo ($\rho(r) \propto r^{-4}$) structure. However the radius of the core in the bulge of model S2 is one third of that in model S3, and the central density of model S2 is larger than that of model S3. Makino, Akiyama & Sugimoto (1990) showed that the surface brightness distribution of a stellar system with density distribution proportional to $r^{-4}$ in the outer region and $r^{-2}$ in the inner region can be fit by any de Vaucouleurs type $r^{1/m}$ law (de Vaucouleurs 1948).

## 5. Discussion

### 5.1. *Summary*

We have shown that starbursts in galaxies with an accreting gas halo can produce a stellar shell with a mass of several $10^{10} M_\odot$ and a size of several kpc by using a similarity solution and a condition for gravitational instability of a gaseous shell (Section 2). In order to investigate the evolution of the shell and its dynamical effects on the structure of the host galaxy, we performed three dimensional $N$-body simulations. The initial stellar shells were simply assumed to be thin and spherical, and were evolved in two case of disk models: a fixed disk potential and a live stellar disk. We used the latter model to calculate the structural evolution of the disk caused by the gravitational interaction of the shell and disk stars. The main parameters are the initial velocity dispersion of the shell stars and the mass of the shell. Our numerical results are summarized as follows.

1. The relaxation process and resultant structure are significantly affected by the external potential, and also depend on the velocity dispersion of the initial shell stars as well as the mass of the shell. We found two new types of relaxation which have not been known in the previous collapse simulation of collisionless system without external disk potential; (a) Relaxation is mainly due to phase-mixing caused by the disk potential. During the relaxation multiple-cylinder structures are formed, which finally relax to a vague bulge in $10^9$ yr. (b) Relaxation is dominated by self-gravity of the shell under the influence of the disk potential. An expanding torus appears and a few clumps are formed in this torus near the disk plane. Finally, a compact high density core with a diffuse halo is formed by merging of these clumps.

2. During the relaxation of the shell, the disk thickens. Non-axisymmetric structure (such as spirals) disappears due to the heating of disk stars by the gravitational interaction between the disk and shell. This heating is more effective in dynamically cold shell models, in which an oval density wave expands toward the outer region. In the relaxation process of the dynamically cold shell, a few clumps grow and heat up disk stars.

3. Angular momentum of the disk is transferred to the shell stars during relaxation by dynamical friction between the clumps consisting of shell stars and disk stars. As a result, the final bulge begins to rotate, although the shell initially had no rotation. Mean rotational velocity of the bulge is $\sim 50$ km s$^{-1}$ at 1 kpc, and the spread of the line-of-site velocity in their position-velocity map decreases with the radius. These results are consistent with recent observations of the bulge of our Galaxy (Nakada et al. 1993; Minniti et al. 1992).

*5.2. The origin of S0 galaxies*

Formation of S0 galaxies has been a controversial problem. There appears to exist two scenarios for the formation of S0 galaxies. (A) The properties of S0, such as large bulge/disk ratio, were determined by the condition at galaxy formation. According to Larson (1976), the bulge/disk ratio depends mostly on a star formation rate during initial collapse of a proto-galaxy and how it varies with time: The spherical component is formed in an early phase of rapid formation of stars. In this mechanism, the large bulge/disk ratio of S0 galaxies is caused by a high star formation rate in a high density proto-galactic core. (B) The large bulge/disk ratio was produced by the interaction between galaxies and their environment at the early stage of galaxy formation: This scenario includes stripping of gaseous envelope of gas-rich galaxies at collapse of a cluster of galaxies (Larson, Tinsley, Caldwell 1980), merging of proto-galaxies (Silk, Norman 1981), accretion of small satellites (Pfenniger 1993) or starbursts induced by galaxy-galaxy interactions (Sofue, Habe 1992).

We also suggest that the scenario of bulge formation following a starburst-induced stellar shell and its relaxation is particularly important for producing a large bulge/disk ratio. Our simulations end up of looking like S0 galaxies (e.g. de Carvalho, da Costa 1984; Hamabe, Wakamatsu 1989). Figures 11(a) and (b) are density contours of disk component projected onto the x-z plane for model S2 and S3 at $t = 2.0$, respectively. Figures 12(a) and (b) are the same, but for both disk and bulge components. These figures show that, while the disk components of the two models are quite different, the composite isophotes resemble those of S0 galaxies. In the observational sequence S0a – S0c (van den Bergh 1976) the bulge is most compact in S0c galaxies. Our models indicate that compact bulges are associated with thick disks, and vice versa. Namely, they suggest that the more massive is the bulge compared to the disk, the thinner is the disk compared to the disk radius. In fact, we have found that the observed disk-to-bulge mass ratio appears to be inversely proportional to the scale thickness-to- scale radius ratio of the disk (Wada, Habe, Sofue, Taniguchi, in preparation).

Our model of S0 formation is consistent with the key observational properties. There is no spiral in the disk because of interaction with the shell stars. A hot X-ray halo (Canizares, Fabbiano, Trinchieri 1987; Awaki et al.1991) originates in the superwind phase (Tomisaka, Ikeuchi 1988). If their structure is due to a process triggered by galaxy-galaxy interactions, the presence of S0s would increase with local number density of the galaxies, which is indeed observed (the morphology density relation: Dressler 1980). A galaxy evolves into a gas-poor disk system, as the gas is used up in the starburst phase.

We simply assumed an initially thin stellar disk with no spheroidal stellar component for the live disk model in section 4.2. This model is based upon Larson's model for forming disk galaxy with a small bulge/disk ratio (Larson 1976). This disk dominant galaxy can be formed from the collapse of a rotating proto-galactic gas cloud before star formation. Even if a small bulge already existed before the stellar shell formed, our results would not change significantly. A shell with small velocity dispersion would continue to be strongly affected by the disk potential away from the central region. Interaction between a shell and a disk with a massive pre-existed spheroidal component would be an interesting subject.

The initial velocity dispersion of the shell stars is one of important parameters for dynamical evolution of the system. It would be generated by the Rayleigh-Taylor instability of the gas shell, and by the time difference of star formation in the decelerated gas shell, in addition to the star formation process itself (i.e. gravitational and the thermal instability of the gas shell). However, it is hard to determine the initial velocity dispersion of the shell stars either observationally or theoretically. In order to investigate this process theoretically, two or three dimensional hydrodynamical simulations of shell formation by superwind taking into account self-gravity and star formation of gas would be necessary.

There is a possibility that newly formed stars exist far above a disk of nearby starburst galaxies. For example, the central region of the typical starburst galaxy M82 is associated with a large-scale outflow of dense molecular gas (Nakai et al 1987). It would be expected that this molecular cylinder is a site of star formation. In fact, ionized gaseous filaments are associated with dark filaments of M82 (Lynds and Sandage 1963), and UV radiation from newly born OB stars would be a possible excitation source. Detailed observations of the starburst regions will bring us important information about kinematics of the newly formed stars outside the disk.

Some authors proposed that the bulge of some galaxies may be accumulated from the merger of satellites, such as globular clusters or dwarf irregular galaxies (Schweizer and Seitzer 1988). From a dynamical point of view, this model is basically the same as the collapse of the shell stars in our model. However, there is a serious problem in this scenario. Stars in dwarf galaxies are metal poorer compared to more massive galaxies. This is because the low star formation rate in dwarf galaxies due to their low ISM density, and mass of the dwarf galaxies is so small that they cannot confine metal enriched gases ejected from stars. On the other hand, bulge stars in our Galaxy are metal rich (Rich 1988). Chemical evolution model of the bulge of our Galaxy suggests that the bulge was formed in the initial $10^9$ yr; the IMF should have been more gentle than the Salpeter's IMF; and the time scale of star formation should be less than $10^8$ yr (Arimoto and Yoshii 1987). This means that bulge stars were made from gases contaminated by gas originated in the primeval starbursts. Population synthesis simulations have also shown that a star formation history in a bulge is the same as that in the bulge of our Galaxy (Jablonka, Arimoto 1992). Consequently, recent merger events cannot produce the bulge of our Galaxy. On the other hand, in our starburst model, the gases in the initial gas shell are mixed with metal rich superwind caused by the nuclear starburst, and it is natural that the bulge stars formed from the shell stars are already metal rich.

*Acknowledgments*

We are grateful to Professors D. Sugimoto, T. Ebisuzaki, and J. Makino for letting us use their *GRAPE-3*. We also acknowledge Dr D. Walsh for reading the manuscript and making helpful suggestions, Professor S. Sakashita for his continuous encouragement, and Dr P. Teuben for his useful graphic program for *N*-body simulations. This work was partly supported by the Grant-in-Aid for Scientific Research (C) (04640261) of the Japanese Ministry of Education, Science and Culture.

**Figure captions**

**Figure 1.** Evolution for model C projected onto the x-z plane. Time is shown in the upper right-hand corner in unit of $10^8$yr. All frames are 20 kpc × 20 kpc.

**Figure 2.** Evolution for model B projected onto the x-y plane (left panels) and onto the y-z plane (right panels). All frame are 20 kpc × 20 kpc.

**Figure 3.** Evolution for model $A_2$ projected onto the x-y plane (left panels) and onto the x-z plane (right panels). All frames are 20 kpc × 20 kpc.

**Figure 4.** $M_s/M_d - \sigma_s^2$ diagram, where $M_s$, $M_d$, and $\sigma_s$ are masses of the shell and the disk, and the initial velocity dispersion of the shell stars, respectively. The two horizontal lines, $\sigma_d$, derived from the z-component of the Jeans equation for a Miyamoto - Nagai potential disk at $R = 1.0$ and at $R = 10.0$. Unit of velocity is 10 km s$^{-1}$. Relaxation types of models $A_1$, $A_2$, $A_3$, B, and C are classified into Type I, II, and III (see Section 4.1).

**Figure 5.** Evolution of model S2 projected onto the x-y plane for its disk stars only (*center*), for shell stars only (*left*), and for both shell and disk stars on the x-z plane (*right*). Times are shown in the upper right-hand corner of the right panels. All frames are 24 kpc × 24 kpc.

**Figure 6.** Same as Fig.5, but for model S3.

**Figure 7.** Time variation of mean scale-height of the disks of models S1, S2, S3, NS. Unit of the scale-height is kpc, and that of time is $10^8$ yr. Model NS is a comparison model as described in Section 4.2, in which the shell was added to the stellar disk at time = 0. The open circles represent a scale height of a disk without the shell.

**Figure 8.** Evolutions of mean vertical velocity dispersion of disk stars for models S2 and S3.

**Figure 9.** Time variation of specific angular momentum normalized by the initial value around the z-axis for disk stars (left vertical axis) and for shell stars (right vertical axis) of models S1, S2 and S3.

**Figure 10.** Density profiles of the final shell stars for models S2 and S3. Lines of $\rho \propto r^{-2}$ and $\rho \propto r^{-4}$ are also shown. Both *bulges* have a core ($\rho \propto r^{-2}$)-halo ($\rho \propto r^{-4}$) structure. The dynamically colder shell model(S2) has a smaller size and higher density core, comparing to S3 model.

**Figure 11.** (a) Surface density contours of edge-on view only for the disk stars of model S2. (b) Same as (a) but for model S3. Size of the frames are 24 kpc.

**Figure 12.** Same as Fig.12, but for both disk and bulge stars.